# Charge-based computing with analogue reconfigurable gates

Alexantrou Serb, Ali Khiat, Themis Prodromakis

**As the world enters the age of ubiquitous computing, the need for reconfigurable hardware operating close to the fundamental limits of energy consumption becomes increasingly pressing. Simultaneously, scaling-driven performance improvements within the framework of traditional analogue and digital design become progressively more restricted by fundamental physical constraints. Thus, a true paradigm shift in electronics design is required for fuelling the next big burst in technology. Here we lay the foundations of a new design paradigm that fuses analogue and digital thinking by combining digital electronics with memristive devices for achieving charge-based computation; information processing where every dissipated charge counts. This is realised by introducing memristive devices into standard logic gates, thus rendering them reconfigurable and able to perform analogue computation at a power cost close to digital. The power of this concept is then showcased by experimentally demonstrating a hardware data clusterer and a fuzzy NAND gate using this principle.**

Realising the rapid expansion of the Internet of Things (IoT) relies on the availability of energy, area and computationally efficient yet affordable and often-reconfigurable hardware platforms that could allow for bespoke customisation[1]. Within the fully digital Von Neumann design paradigm that still dominates modern electronics, downscaling of integrated circuits[2] has been the main driver for lowering power dissipation; a process now reaching its physical limits. Simultaneously, reconfigurability has remained mainly in the domain of software engineering, relying for its physical implementation on dedicated memory blocks and progressively bottlenecked by power-hungry data transfers between physically separate memory and processing elements[3]. The rich landscape of modern electronics design became even more diverse with the steady introduction of memristive devices[4] into the family of standard electronic components[5,6]. The ability of memristors to act as electrically tuneable multi-level[7], non-volatile resistive loads[8], combined with their inherently scaling-friendly[9,10], low power[11] and back-end integrable[12] fabrication processes has rendered them a highly promising candidate for use in future electronics applications[13–16]. These properties promote memristors as ideal candidates for achieving in-silico reconfigurability in a post-Moore and post-Von Neumann context, i.e. without relying on front-end integration density for performance and operating on the principle of separate, dedicated memory and processing elements.

In this work, we lay the foundations of a novel design paradigm that amalgamates the analogue non-volatile memory capacity of metal-oxide memristors with the fundamental building blocks of digital design: logic gates. This amalgamation occurs at a fundamental component level, enmeshing memristors and transistors in order to achieve collocation of memory and computation. Our approach is thus distinct from conventional mixed-signal design, whereby the analogue/digital paradigms co-exist and yet remain separate entities, interacting purely at the signal level. This true fusion of paradigms engenders a new set of fundamental building blocks: analogue reconfigurable gates featuring embedded memory. We first prove the competitive advantage of the proposed building blocks that allow computation close to the fundamental limits of energy dissipation. We then demonstrate the reconfigurability modes of a memristor-enhanced inverter by delineating how tuning the memory states of individual devices enables the control of the transfer characteristics of the gate. The proposed design paradigm is completed by introducing appropriate read-out circuits that make our modified gates interoperable to standard digital gates. We envision this new concept becoming a staple in numerous emerging applications and showcase the versatility of the proposed paradigm by experimentally demonstrating two applications: a hardware fuzzy gate and an analogue domain template matcher.

## RESULTS:

### Reconfigurable analogue gate concept, operation and performance

Much like multi-valued logic is a generalisation of standard Boolean logic, the proposed analogue gates are inspired as generalisations of standard logic gates. This is realised through the topology shown in Figure 1a for the analogue inverter (comparison with standard Boolean inverter in Supplementary Figure 1). Every current path to/from the output node of the gate is regulated by the presence of a tuneable resistive element, in our case a metal-oxide memristor (see Supplementary note 1 for the fully general approach). This architecture has the same inputs and outputs as a standard gate but receives analogue inputs and generates an analogue output. Overall, the proposed analogue inverter serves as a potential divider comprising two transistors and two memristors (2T2R). Depending on the precise levels of the input voltages, the output voltage behaviour may be dominated either by the states of the transistors (standard logic gate operation) or by the memristive components and their interrelations (divider operation). The former is obtained when input voltages are clear binary values, whilst the latter at intermediate levels. This occurs because at each edge of the input voltage range one of the transistors always exhibits a source-drain impedance that is sufficiently high to dominate the entire divider and lead to standard Boolean inverter operation. Simultaneously, at intermediate values of input voltage both transistors are open and the memristive potential divider becomes dominant. This introduces a plateau in the transfer characteristic of the inverter, visible in Figure 1d, which controls the shape of the mapping between input and output voltages whilst maintaining the fundamental inverter nature of the circuit (0 maps to 1 and vice versa). Controlling the resistive states of $R_{UP}$ and $R_{DN}$ allows this soft mapping to be reconfigured, with the reconfiguration quality defined by the characteristics of the employed memristive technology.

A key feature of the proposed design is its power efficiency that can be illustrated by comparing an analogue vs a digital inverter. For any given input voltage both inverters can be described to some approximation as two component potential dividers, illustrated in Supplementary Figure 2b. Whenever the input voltage changes from some value $V_{IN,1}$ to some other value $V_{IN,2}$ the corresponding outputs must change from some $V_{OUT,1}$ to some $V_{OUT,2}$; an operation that requires changing the amount of charge stored on capacitor $C_{out}$ through a capacitor current $i_{cap}$ while keeping leakage current $i_{leak}$ low (illustrated in Supplementary Figure 2 with full derivation in Supplementary note 2). In the case of a standard inverter the only possible input (output) voltages are GND and VDD for logic 0 and 1 respectively, which guarantees that one of the transistors M1 and M2 will always be OFF, i.e. at very high source-drain impedance. This, in turn, minimises $i_{leak}$ allowing the inverter to operate close to the well-known theoretical minimum energy $C_{out}\frac{VDD^2}{2}$ for each state change[17]. The analogue inverter is governed by the same fundamental dynamics with the exception that M1 and M2 may both be partially ON at the same time. As a result, the inverter becomes capable of performing analogue-in/analogue-out computation at a fraction of the energy consumption of its digital counterpart (e.g. factor of ~4-5 see Supplementary note 3 for details).

Given any fixed set of resistive states for $R_{UP}$ and $R_{DN}$ an analogue gate will implement a specific soft input/output mapping, with each memristor constituting a design degree of freedom (dof). Since every memristor augments the impedance seen from the output node to either supply or ground along a unique current path, these design degrees of freedom are linearly independent. Thus, the proposed analogue inverter features two degrees of reconfigurability freedom, whose span is bounded by the range of resistive state values that the corresponding memristor can attain. The design of the non-memristive part of the system, and in particular the settings of the key design parameters of transistor aspect ratio $\frac{W}{L}$ and power supply voltage VDD, may then be tailored so as to optimise gate functionality given the chosen memristor technology's inherent resistive state ranges (See Supplementary Figure 3 and Supplementary Table 1). The 2-dof reconfigurability space of the inverter is illustrated in Figure 1b, showing two useful, orthogonal soft inverter mapping control modalities. In both cases, the values of the two memristors $R_{UP}, R_{DN}$ are altered simultaneously first under the constraints $\frac{R_{UP}}{R_{DN}} = c$ (ratio-fixed modality) and then under the constraints $R_{UP} + R_{DN} = c$ (sum-fixed modality), where $c$ is a suitable constant in each case. The two control modalities exert orthogonal effects on the plateau in the inverter's transfer characteristic (see Figure 1c). The ratio-fixed modality controls the breadth of the plateau by altering the balance between the total memristor-transistor (source-drain) impedance whilst the sum-fixed modality controls the altitude of the plateau by altering the voltage distribution within the memristive potential divider. Provided that the distribution of voltage between memristors is not affected by the overall voltage drop across them, these modalities do not interact with each other (see Supplementary Figure 4). Figure 1c presents experimental evidence of an analogue inverter whose memristors have been successively programmed into four configurations: $\{R_{UP}, R_{DN}\} \in \{HH, HL, LH, LL\}$ where $H, L$ stand for high and low resistive state respectively. Results demonstrate operation under an ON/OFF resistive state ratio of approximately 10, leading to significant changes in the input/output transfer characteristic of the inverter. The HH-LL pair illustrates modulation of plateau width independent of altitude whilst the HL-LH pair illustrates altitude modulation sans width modulation.

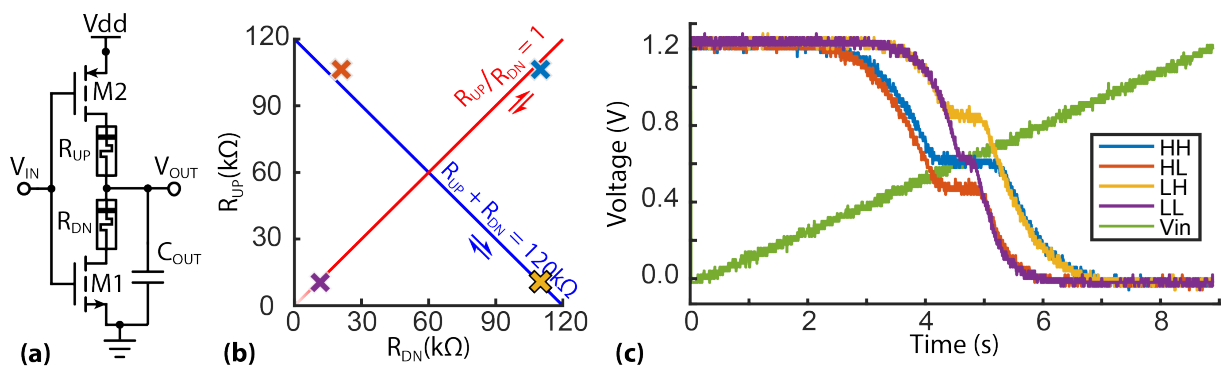

Figure 1: Reconfigurability modalities in analogue inverter gate. (a) Memristor-enhanced analogue inverter topology. (b) Changing the resistive states of the memristors $R_{UP}, R_{DN}$ in the inverter so as to keep their sum (along blue line) or ratio (along red line) constant offers flexibility in controlling the inverter's transfer characteristics. The constant sum modality allows independent control of transfer characteristic's plateau height whilst the constant ratio modality allows for independent control of the plateau's width (see Supplementary figure 4). Colour-coded crosses correspond to the $R_{UP}, R_{DN}$ configurations used in the results of panel (c) (see Supplementary Table 2 for details). (c) Four, measured examples of analogue inverter transfer characteristics corresponding to the cases where $R_{UP}$ and $R_{DN}$ are both high (HH), high and low (HL), low and high (LH) and both low (LL) respectively. The measured input voltage during the HH trial is shown in green as $V_{in}$ (similar for all trials). Note independent modulation of plateau width and altitude by the sum and ratio between $R_{UP}, R_{DN}$.

**Interoperability with standard digital electronics**

No electronic system can become commercially competitive vis-a-vis standard CMOS technology if it cannot be both read in a simple and efficient manner and modularly chained, which in our case means that the output of an analogue gate has to be a suitable input for the next one. Since the proposed gates employ analogue voltages as both inputs and outputs this compatibility is ensured. This is further illustrated in Supplementary Figure 5 where a NAND gate operates on the basis of input received from an analogue inverter. Transferring from analogue logic to Boolean and/or brain-inspired can be easily achieved using a read-out circuit consisting of a simple inverter fed through a mirror supply as shown in Figure 2. The analogue-to-Boolean link rests on the fact that the read-out inverter will be characterised by a switch-point voltage, i.e. an input voltage level at which both transistors are simultaneously ON and the inverter output voltage is close to the middle of the supply. Any input voltages above switch-point will be digitised to 0 whilst values below switch-point will digitise to 1 (Figure 2, $V_{OUT1}$). A small range of input values very close to switch-point, however, will lead to an unclear digitisation that may stochastically result in digital 1 or 0 values. Notably, changing the resistive states of the memristors in any analogue gate can alter the input variable space regions which lie above/below the read-out inverter switch-point and thus indirectly tune the overall mapping from analogue gate inputs to digital output. The switch-point of the read-out inverter is determined by the aspect ratio (W/L) of its constituent transistors and is a design parameter; though there is no reason memristors cannot be used to render the read-out inverter reconfigurable too. On the other hand, converting Boolean-to-analogue requires no conversion as any Boolean input level is automatically a valid analogue input.

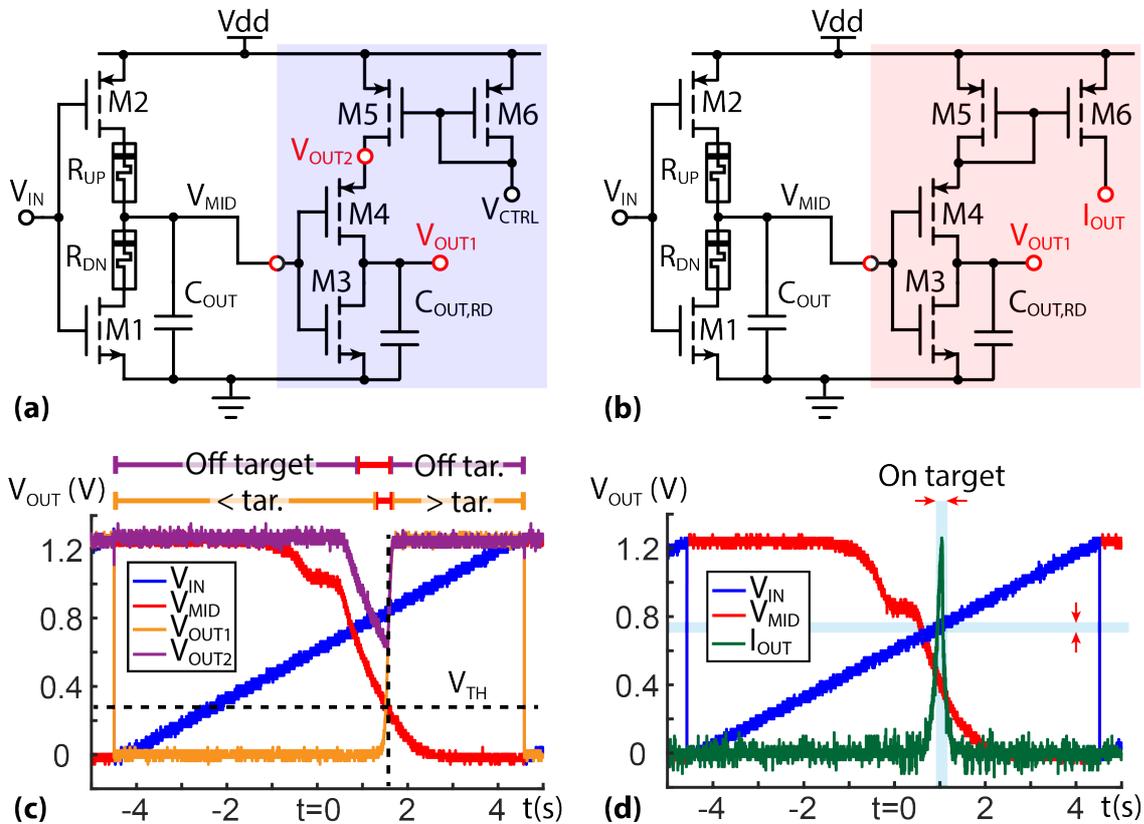

**Figure 2: Reading analogue gates.** (a) Read-out circuit designed to receive an analogue gate output ($V_{MID}$) and then digitise to 1 or 0 based on whether $V_{IN}$ is above or below some threshold $V_{TH}$ (at $V_{OUT1}$) and also indicate whether $V_{IN}$ is close or far from $V_{TH}$ (at $V_{OUT2}$). (b) Alternative read-out circuit where $V_{OUT2}$ has been replaced by $I_{OUT}$, which now indicates proximity of $V_{IN}$ with $V_{TH}$ by sourcing a large current only if $V_{MID} \cong V_{TH}$. (c) Measured results from the read-

out circuit in (a). $V_{OUT1}$ successfully digitises the analogue inverter output $V_{MID}$ through most of its input voltage range ($V_{IN}$) as shown by the orange bar above the plot (red segment indicates potentially ambiguous digitisation). $V_{OUT2}$, on the other hand exhibits a dip only when $V_{MID}$ is sufficiently close to $V_{TH}$, directly indicating whether $V_{IN}$ is on-target (i.e. close to $V_{TH}$) or off-target. This is illustrated by the purple bar above the plot. (d) Measured results from the read-out circuit in (b). Only $I_{OUT}$ shown for clarity. $I_{OUT}$ peaks within a narrow range of input voltages satisfying $V_{MID} \cong V_{TH}$ (on target).

The mirror supply shown in Figure 2a is not strictly necessary for the digitisation strategy described above, but offers an interesting alternative approach. The principle of operation relies on the observation that the read-out inverter conducts most current at the switch-point voltage, when both its transistors are maximally ON simultaneously and digitisation at node $V_{OUT1}$ is unclear. The mirror supply exploits this by attempting to force a reference current into the inverter. If the reference current is chosen appropriately, the voltage on node $V_{OUT2}$ from Figure 2a is driven towards a digital 0 only when the inverter is sufficiently close to its maximally conducting state (both transistors simultaneously open). The circuit therefore offers a way of performing digitisation by mapping a very specific analogue gate output level to a digital 1, i.e. allowing that particular value to act as a target and the analogue gate to output an off-target/on-target response (see 2c, purple bar). Modifying the magnitude of the reference current will tune the analogue gate output level range for which digitisation using this strategy will return a digital 1. A slight modification of the mirror supply leads to the circuit depicted in Figure 2b, which mirrors the read-out inverter current directly to a circuit terminal. Providing the digitisation output in the form of a current allows for easy summation of digitisation results from many analogue gates. The resulting, summed current can then easily be provided as input to an integrate and fire neuron as used in neuromorphic engineering[18]. Notably, both variants in Figure indicate that the on-target range of the input voltage is of the order of 100mV as evidenced by $V_{OUT1}$ voltages that are no longer a clear digital 1 or 0 (Figure a,c) or $I_{OUT}$ currents significantly above baseline (Figure 2b,d). The on-target range will depend on the resistive state values of the memristors as is immediately evident by observing that the traces in Figure 1c will intercept the threshold shown in Figure 2c at different points along the x-axis (and consequently different input voltage levels).

**Case studies: the fuzzy gate and the texel**

The proposed design paradigm forms a generalised framework that can be used to develop a broad range of applications. Here, we showcase two cases, namely: fuzzy gates and template matching. Fuzzy logic relies on analogue-in/analogue-out fuzzy gates in order to perform computation on fuzzy sets and soft versions of standard digital gates act as fuzzy counterparts. These implement a fuzzy input/output mapping that may not necessarily correspond to any of the standard fuzzy logic rules (e.g. Zadeh rules), but can nevertheless be used for function approximation. The example of a fuzzy NAND gate is shown in Figure 3. A fuzzy NAND can be implemented via a three-way divider consisting of four transistors and three memristors (4T3R). Figure 3b shows the transfer characteristics from the two inputs to the output, which now define a surface. Notably, when input A is fully ON (digital 1), then the fuzzy NAND reduces to a fuzzy inverter from B to the output with a mapping determined solely by memristors $M_B$ and $M_C$. We shall term this the fuzzy inverter reduction of B. The same holds when input B is a digital 1. In the case where either of the inputs is fully OFF (digital 0), the output of the fuzzy NAND will always be a digital 1. Changing the resistive states of the memristors controls the shape of the fuzzy function surface whilst retaining its inherent NAND nature. The measured results shown on Figure 3b denote an essentially multiplicative interaction between the fuzzy inverter reductions of A and B, suggesting that the fuzzy function surface may be controlled in a reasonably orthogonal way by varying $M_A$ and $M_B$ at the cost of restricting the possible shapes it may assume. Similar conclusions may be drawn for other types of gates (see Supplementary Figure 6), although fuzzy NANDs are already functionally complete (in the sense that using multiple fuzzy NANDs, mappings corresponding to any other fuzzy gate can be constructed).

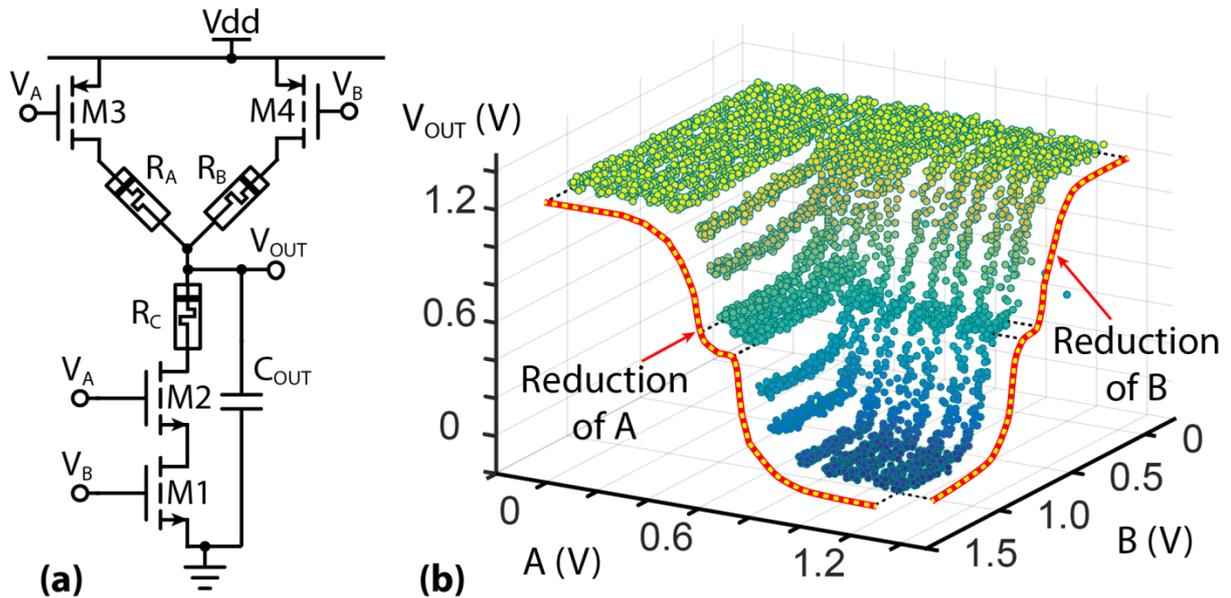

Figure 3: Fuzzy NAND architecture and basic behaviour. (a) Fuzzy NAND topology. (b) Measured fuzzy NAND transfer characteristics. The inverter reductions of A and B (see text) are shown as red/yellow lines. The overall transfer characteristic arises from an essentially multiplicative relation between the reductions of A and B.

Template matching is a technique whereby a small part of a signal (audio/electrical waveform snippet or even image segment) is compared against a stored template. It is one of the most popular approaches for performing neural spike sorting for electrophysiological studies[19] whereby electrical waveforms recorded from neural cell assemblies are template-matched in short snippets of typically 10-20 samples[20]. Its strength stems from the fact that whenever a match is found the system registers the occurrence of a spike and the matching template ID, thus simultaneously providing spike timing and identification information. This concept can be realised via the circuit topology shown in Figure 4a; a simplified version of the analogue inverter with the read-out circuit from Figure 2b. We refer to this circuit as a 'texel' (template pixel) and its operation can be understood as searching for an input voltage value $V_{IN}$ that matches stored value determined by the resistive state of memristor R1. This process is shown in Figure 4b-d by illustrating measured transfer characteristics of a discrete texel circuit whilst results from a 4-texel array are shown in Figure 4e. The array is fed with nine neural spike waveforms from the same database[21] (Supplementary Figure 7) and summing the current outputs of each texel down a common load resistor, as shown in Figure 4e,f. Three spike instances were chosen from each available class of spikes: a low (L), a medium (M) and a high (H) instance corresponding to spikes exhibiting lower than, similar to or higher than class-average voltage levels (see methods and Supplementary Figure 7). The voltage level at the system's $V_{OUT}$ terminal is linked to the degree of matching between the input vector k and the stored template and was directly used as a matching degree metric. Due to the similarity between the H instance of class 1 and the L instance of class 2 and the limited resolution of our instrumentation, the experiment for these two instances was ran only once with a common input vector k (see Supplementary Table 3). Results in Figure 4g show a texel array set up to discriminate for class 2 spikes. Even using only four samples from each waveform (marked in Figure 4e) strong discrimination between templates is clearly achieved. The memristor resistive states were confirmed to remain stable before and after the experiment (Supplementary Table 4).

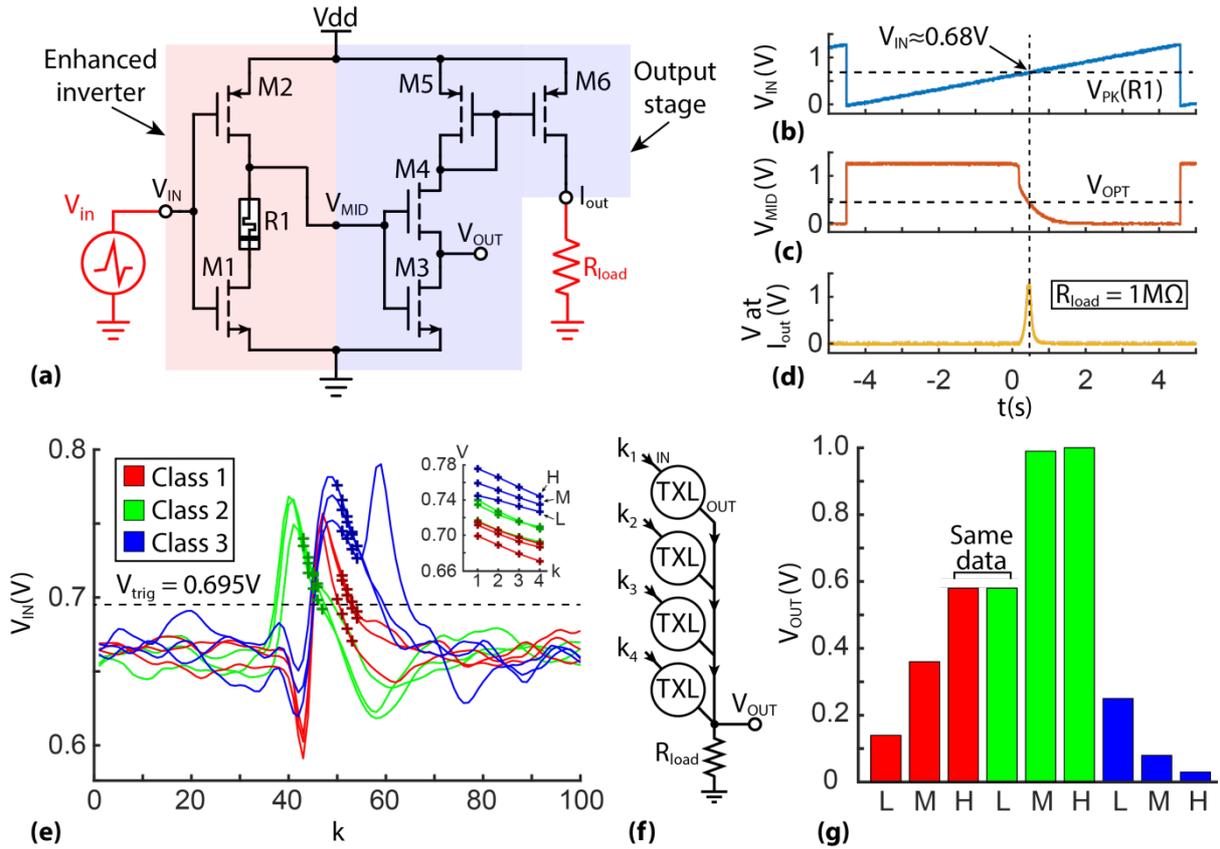

Figure 4: Analogue domain template matching using memristive technologies. (a) Schematic of 'texel' circuit illustrating consisting of an analogue inverter and the read-out stage. (b-c) Texel transfer characteristics from input voltage (b), through mid-point voltage $V_{MID}$ (c) and voltage at the output node (c) marking $V_{IN}$ input voltage level ($V_{PK}$) and $V_{MID}$ voltage level ($V_{OPT}$) at which the output stage sources its maximum current. (e) Selected spike waveforms used as input to the test texel array. Crosses indicate the sample points used to feed the array. k: sample number. $V_{trig}$: texel array sampling trigger level (see methods section). Inset: close-up of the chosen sample points. L, M, H and arrows: Low, medium and high voltage instances of spikes in class 3. (f) Schematic of 4-texel array used to carry out experiments. (g) Measured output voltage when spike samples from (e) are applied to the texel array in (f). L, M, H versions of spikes in each class shown. Higher $V_{OUT}$ voltage means greater degree of matching between input data and stored template. The texel array was programmed to respond best to class 2 spikes. Colours as in (e). Class 1-H and class 2-L results refer to the same experiment (see methods section).

## DISCUSSION:

Our proposed design paradigm must be understood as a simple, powerful and generic tool for truly fusing the analogue and digital ways of thinking. Importantly, the transistors and memristors in Figure 1a can be any elements that modulate a resistance, not even necessarily electrical (could be hydraulic for instance). In every case, the basic concept remains the same: signal-controlled tuneable resistance elements intended to operate as ON/OFF switches are combined with 'initialise & operate' continuously tuneable resistance elements. This combination results in systems where digital and analogue behaviours coexist, with the ON/OFF and continuously tuneable components tasked to sustain each type of operation respectively. This realisation hints towards the potential that the 'division of labour' between analogue and digital paradigms may be possible in many shapes and forms and at many different levels, ranging from the subsystem-level collaboration common in mixed signal designs to the fundamental component level fusion exemplified in this work. In conclusion, this work lays the conceptual foundations and provides the first experimental proof of a new design paradigm for electronics formed by marrying standard, digital circuitry with analogue electronics and rapidly emerging memristive devices to offer analogue computation at (close to) the power/area price of digital with the added benefit of reconfigurability. This is achieved by collocating and enmeshing memory and computation whilst maintaining full interfacing compatibility both with standard digital CMOS and internally (ability to chain analogue gates as is done with digital gates). Finally, the versatility of this new paradigm is illustrated by two independent applications in fuzzy logic and template matching for bio-signals that given the power constraints of alternative technologies would have been impossible to realise.

## Methods

**Memristive device fabrication and specification:** Memristive device fabrication and specification: All the memristors used in the experimental setups are in 3x3 mm$^2$ chips that are wire-bonded to PLCC68 packages. Each memristor is a 20x20 μm$^2$ cross-point of top and bottom electrodes (TE & BE). The BEs were first fabricated on 6 inch Silicon wafer that was thermally oxidised to grow 200 nm thick $SiO_2$, which serves as insulating base layer. Using conventional optical lithography the BEs were patterned with negative-tone resist AZ-2070 followed by a low power and short reactive ion etching (descum) to remove any residual resist on the patterned areas. Then, 5 nm Titanium (Ti) and 10 nm Platinum (Pt) film were deposited with electron beam evaporation at low rate (o.5 Å/s), with Ti serving for adhesion purposes. Leybold_Lab700eb tool that has high crucible-wafer distance (> 1m) was used for evaporation to guarantee parallel deposition. The combination of parallel evaporation and negative-tone resist which has undercuts after development ensures a good liftoff process, resulting in well-defined electrodes without wings (fences) that would affect the subsequent thin layer, thus becomes harmful for the final device. Next, similar photolithography was carried out using the active-layer mask that allows depositing the metal-oxide active bi-layer everywhere except on the BE pads, followed by 1 min descum. Lambda controlled plasma assisted reactive magnetron sputtering (Leybold Helios Pro XL) was used to deposit the active layers; 25 nm $TiO_2$ followed by 4 nm $Al_2O_3$, at room temperature. $TiO_2$ was sputtered from Ti metal target with 8 sccm $O_2$, 35 sccm Ar flows and 2 kW at the cathode, and 15 sccm $O_2$ flow and 2 kW at an additional plasma source. $Al_2O_3$ was sputtered from Al metal target with 15 sccm Ar flow and 100 W at the cathode, and 25 sccm $O_2$ flow and 1.5 kW at the additional plasma source. Before any sputtering, the additional plasma source was used for an extra clean of the substrate with 8sccm $O_2$, 10 sccm Ar and 2 kW. Optiwet-ST30 tool was then used to perform liftoff process with the following parameters; 3 mbar pressure and 60 °C hot NMP for 1 hour. This tool ensures 100 % liftoff yield even after sputtering which deposits material everywhere making the liftoff difficult, and even at large features (pad areas) that tends to stick to the surface after dissolving the resist. 1 min descum is needed to clean the surface before TE lithography. The 10 nm Pt TEs were patterned and defined in similar fashion to the BEs. No sonication was used in this process. Finally, 3x3 mm$^2$ chips were diced for wire-bonding.

**Experimental set-ups and procedures:** All experiments illustrated in Figures 1-4 were carried out on circuits prototyped on breadboard or strip-board. External power supplies and signal generators were used to supply both signal inputs and power whilst results were gathered exclusively by oscilloscope. For these experiments packaged devices were used, connected to the set-ups via breakout boards. This is significant because it demonstrates functionality after wafer dicing and wire-bonding. All experiments were ran under a power supply between 1.2-1.3V (single decimal figure precision power supply). In each case, the memristive devices used were placed in the required resistive states using an ArC ONE instrumentation board (ArC instruments, UK). All devices used for all experiments were located on the same die, i.e. only one memristive device package containing a total of 32 memristors was sufficient to carry out all the work presented here.

Additional notes on fuzzy NAND experiment: Results for the fuzzy NAND experiment were taken 'strip-by-strip' by setting input A to a succession of fixed values and sweeping input B for each of those values by use of ramp (saw-tooth) signals. Input A was stepped with 100mV resolution throughout the entire power supply range except between 0.5-0.8V where it was stepped with 50mV resolution for enhanced visibility.

Additional notes on spike sorting application experiment: A 4-point texel array with common load resistance of 300kΩ was implemented. The signals fed into this system arrived from two, dual-channel benchtop power supplies with two significant decimal digits resolution. The benefit of using synthetic neural recording input data is that it contains ground truth information on spike identification and timing. On that basis an automatic sample selection script was ran on each spike instance available in the dataset in order to choose which data-samples from each instance are to be fed into the texel array for matching against a stored template. The script operated as follows: the data-points in each spike instance were read sequentially and once a trigger threshold Vtrig was exceeded for the first time the script skipped six samples and then choose the subsequent four as candidate inputs for the texel array set-up. This methodology was chosen because it rendered the three classes of spikes visibly distinguishable despite the use of only four template points. The overall 'trigger and sample' approach is similar to the work by Restituto-Delgado et al[22]. In a more mature system implementation a larger texel array containing more than four samples would be used. Next, the extracted candidate four-texel sample sets were separated by single unit-template class. From each class, three texel sets were chosen for further processing: one featuring typical (M), one featuring lower than usual (L) and one featuring higher than usual (H) voltage values (selection shown in Supplementary Figure 7). Waveforms where the presence of more than one spike within each instance had corrupted the output of the sample selection script were automatically excluded from the selection. The voltage range of all nine selected sample sets (L, M, H instances for each of the three classes) was then adjusted by application of a common pair of gain and offset settings (Gain: 0.1; Offset: 0.66V). The adjusted texel data-point voltages were then suitable for working with the input voltage values the texel circuits were built to discriminate between. These adjusted values are shown in the inset of Figure 4e and were used as the input to the texel array after being rounded to 10mV precision (two significant decimal digits). This procedure caused the rounded texel voltages of the H instance of class 1 and the L instance of class 2 to completely overlap, hence that experiment was conducted only once for both cases.

**Acknowledgements:** This work has been supported by the Engineering and Physical Sciences Research Council (EPSRC) grants EP/K017829/1 and the EU FP7 RAMP project.

**Author contributions:** AS conceived and ran the experiments, processed the resulting data and wrote the manuscript, AK developed the process flow and fabricated the memristive devices and TP contributed significantly to the organisation of the manuscript and the construction of the abstract. All authors contributed to reviewing and refining the manuscript.

**Competing financial interests:** The authors declare no conflict of interest of any kind.

**Data availability:** TBD

# Supplementary material:

# Supplementary notes:

**Supplementary note 1 – Generalised approach to designing analogue gates using memristors**

The fundamental principle behind our proposed paradigm is illustrated in its most general form in Supplementary Figure 8 for both the inverter and the NAND gate. Each memristor performs a unique function within each topology; modulating drain-source resistance or source-degenerating a transistor or both. Inverter case: $R_A$ and $R_D$ source-degenerate transistors M2 and M1 respectively whilst $R_B$ and $R_C$ modulate their effective drain-source resistances respectively. NAND case: similar to the inverter but $R_F$ simultaneously source-degenerates M2 and modules effective drain-source for M1. It is interesting to note that the full general analogue inverter has therefore 4 degrees of freedom, whilst the fully general analogue NAND features 7 dofs; one short of double the inverter's. due to the shared functionality of $R_F$ in the NAND gate. We therefore conclude that gates that include many instances where one transistor's source connects directly to exactly one transistor's drain will feature relatively fewer degrees of freedom. If, however, the transistors are not connected one source-to-one drain a memristor may be introduced in front of each transistor much like the configuration at the output of the NAND gate (node $V_{OUT}$ in Figure 3a and Supplementary Figure 8).

**Supplementary note 2 - Estimating power dissipation**

A crucial aspect of the proposed design is its power efficiency. This is best illustrated by inspecting the energy dissipation of an ideal digital inverter and its analogue counterpart. The energy required to flip a standard inverter's state $E_{flip}$ depends on the output capacitance $C_{out}$ and the power supply voltage $VDD$ and is given by the well-known formula:

$$E_{flip} = C_{out} \frac{VDD^2}{2} \qquad (1)$$

In order to investigate power dissipation in an analogue, reconfigurable inverter we consider a very simplified circuit where both transistors and memristors are treated as linear resistors that remain constant for any fixed input $V_{IN}$ as illustrated in Supplementary Figure 2b. The objective is to compute the energy cost involved in moving the output voltage of the analogue inverter from its initial state $V_{OUT,1}$ under input voltage $V_{IN,1}$ to a new state $V_{OUT,2}$, as imposed by a new input voltage $V_{IN,2}$. The current leaving the power supply $I_{VDD}$ is given by Kirchhoff's law:

$$I_{VDD}(t) = \frac{VDD - V_{OUT}(t)}{R_1} = \frac{VDD - \Delta V_{OUT} e^{-t/R_\| C_{out}} - V_{OUT,2}}{R_1} = \frac{VDD(1 - Q_{div}) - \Delta V_{OUT} e^{-t/R_\| C_{out}}}{R_1} \qquad (2)$$

where $Q_{div} = \frac{R_2}{R_1 + R_2}$, $\Delta V_{OUT} = V_{OUT,2} - V_{OUT,1}$, $R_\| = \frac{R_1 R_2}{R_1 + R_2}$, $R_1$, $R_2$ the equivalent memristor-transistor resistances at $V_{IN} = V_{IN,2}$ and keeping in mind that $V_{OUT,2} = VDD \cdot Q_{div}$.

Integrating (2) over time for an interval of time $t_{set} \equiv lR_\| C_{out}$ where we consider the system to have satisfactorily converged to its equilibrium value ($V_{OUT,final} \approx V_{OUT,2}$) we obtain total charge usage $Q_{tot}$ of:

$$Q_{tot}(t_{set}) = \int_{t=0}^{t=t_{set}} I_{VDD} dt = t_{set}\frac{VDD}{R_1+R_2} + C_{out}\Delta V_{OUT}Q_{div}(1-e^l) \quad (3)$$

We notice that the first term is a constant leakage down the inverter (leakage term) and it depends on the total inverter impedance and the time necessary for the computation to be concluded to within tolerance. This is illustrated in Supplementary Figure 2b as i_leak. The second term includes the ideal charge transfer required to change the voltage at the output node by $\Delta V_{OUT}$, $Q_{ideal} = C_{out}\Delta V_{OUT}$ (charging term i_cap in Supplementary Figure 2b). The $Q_{div}$ in the charging term is best understood as the extent to which the $C_{out}$ capacitor current flows into the power supply or the ground. With $Q_{div}$ close to zero $C_{out}$ charges/discharges preferentially into the ground (similar to a standard inverter toggling from output 1 to 0 – the output signal transition is achieved primarily by sinking charge from the output capacitance into ground) whilst for $Q_{div}$ close to unity the supply is preferred (standard inverter toggling from output 0 to 1). We also note that the charging term may be positive or negative depending on the relationship between $V_{OUT,2}$ and $V_{OUT,1}$. Without loss of generality we consider the case where $V_{IN,1} > V_{IN,2}$, $V_{OUT,1} < V_{OUT,2}$ and the charging term is positive.

Translating charge into energy dissipation we can compute an upper bound by rounding $Q_{div}$ to unity and considering that every charge $q$ leaving the power supply will (eventually) reach ground dissipating $qVDD$ energy. We thus obtain:

$$E(t) < Q_{tot}(t_{set})VDD < t_{set}\frac{VDD^2}{R_1+R_2} + VDD\,C_{out}\Delta V_{OUT}(1-e^l) \quad (4)$$

This highlights three points: First, the analogue inverter has an (upper bound) energy dissipation given by a charging term that reduces to the standard inverter (within factor of 2) dissipation for $\Delta V_{OUT} = VDD$ and $t \to \infty$ plus a leakage term. Second, the leakage term depends on the in-operando impedance of the inverter whilst the charging term only depends on output capacitance. Third, longer waiting times lead to more accurate computations ($V_{OUT}(t_{set})$ closer to the ideal equilibrium value), but incur a larger leakage energy penalty.

Finally, it can be shown that by expressing $R_1 + R_2$ in terms of $Q_{div}$ and $t_{set}$ in terms of $l$, eq. (3) can also be expressed as:

$$Q_{tot}(t_{set}) = lC_{out}VDD\,Q_{div}(1-Q_{div}) + C_{out}\Delta V_{OUT}Q_{div}(1-e^l) \quad (5)$$

The fact that charge consumption can be expressed as function of $Q_{div}$ only, shows that the absolute values of $R_1$ and $R_2$ are only significant for setting the temporal dynamics (via the $R_{||}C_{out}$ constant – the units of $l$); charge dissipation depends only on their relationship. This is significant because it suggests that analogue gate speed can be traded off against memristor resistive state. Eq. (5) also reveals that for $\Delta V_{OUT} \approx VDD$ and suitable $Q_{div}$ the leakage and charging terms will be broadly comparable even for $l$ values sufficiently long to allow the system to converge to equilibrium (e.g. within 2% for $l = 4$).

In conclusion, the calculations above suggest that analogue computation is achievable at an energy price close to digital under the simplifying assumptions about transistor and memristor resistances made by equations (1-5). Moreover, in practical electronics input voltages can only cross from one level to the other within a finite interval of time. This causes even the standard Boolean inverter to spend some time with both its transistors ON when the input voltage is between digital 1 and 0 with an associated energy cost ignored by eq. (1). The precise impact of finite input transition times on the results in eq. (4) lies outside the scope of this paper. Results on power dissipation are in line with expectations from the simulations on more realistic fuzzy and standard inverters, implemented in a 0.35 micron commercially available technology that are shown in Supplementary note 3.

**Supplementary note 3 - Power estimations for analogue inverter operation**

Estimation of the operating power budget of an analogue inverter was investigated through simulation on the industry-standard Cadence tool (see Supplementary methods). In order to give a very conservative and operationally relevant estimate, a modified version of a full texel circuit, as illustrated in Figure 4a, was simulated (full texel schematic used in Supplementary Figure 9b) within the power dissipation estimation test bench shown in Supplementary Figure 9a. The reference technology was AMS 0.35 micron (C35). Resistors were used to model memristors, and the amount of charge removed from the power supply to carry out the computation was taken as a proxy for power dissipation.

Operating power estimations were benchmarked for the following analogue computation: input voltage rises from 1.55V to 1.7V. These values guaranteed a visible change in the system output voltage level, as evidenced in Supplementary Figure 10. By the time the output voltage stabilises the overall amount of charge removed from the power supply is approximately 46fC. This compares favourably with the ~1.25fC charge dissipated by an industrially-designed minimum size inverter for a single digital state transition (input 0 to 1) in the same technology, as shown in Supplementary Figure 11. Therefore for the charge dissipation price of ~39 inverter toggles the texel carries out an analogue input-output mapping operation. Notably, the energy price includes the operation of two analogue inverters (driving inverter from test bench and inverter included within texel) plus the read-out stage. Furthermore, the texel circuit used in this study was not optimised for low power dissipation but is provided as a working example that can be set up with minimum design effort.

## Supplementary figures:

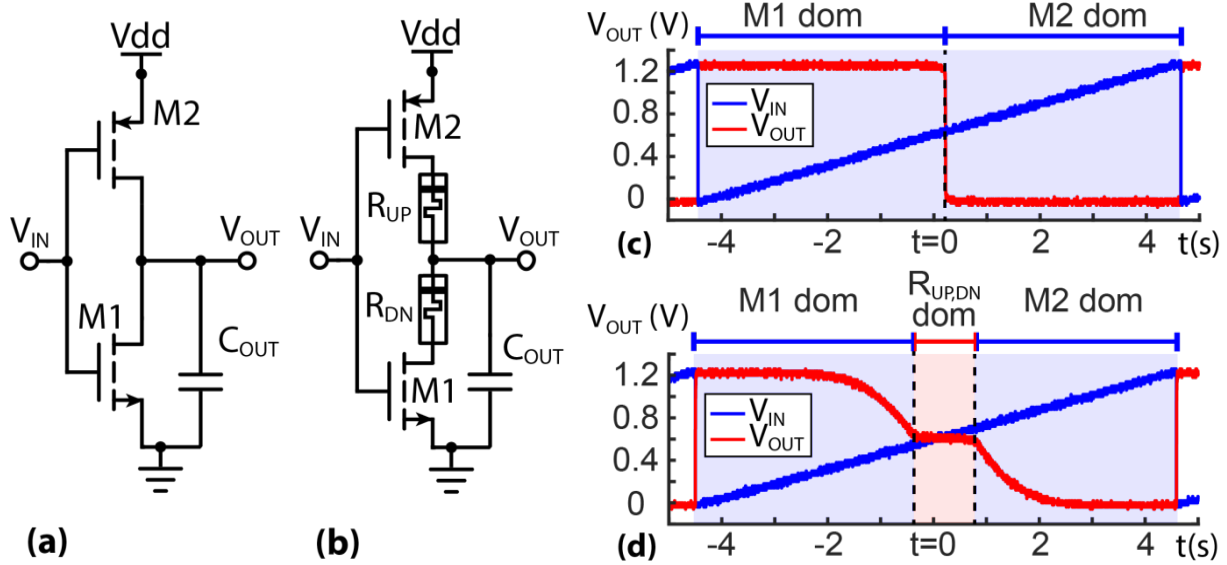

**Supplementary figure 1:** Analogue inverter gate architecture and basic behaviour. (a) Standard Boolean and (b) memristor-enhanced analogue inverter topologies. (c) Measured Boolean and (d) analogue inverter transfer characteristics. Devices $R_{UP}$ and $R_{DN}$ in the analogue inverter are memristors. The skew in the input/output transfer characteristic introduced by the memristors is evident. In both (c) and (d) blue/red bars above the figures indicate which component's resistance dominates the total impedance of the potential divider formed between Vdd and GND. The red-shaded, plateau region in the transfer characteristic of the analogue inverter shows the range where the memristors provide most of the impedance as transistors M1 and M2 are simultaneously open.

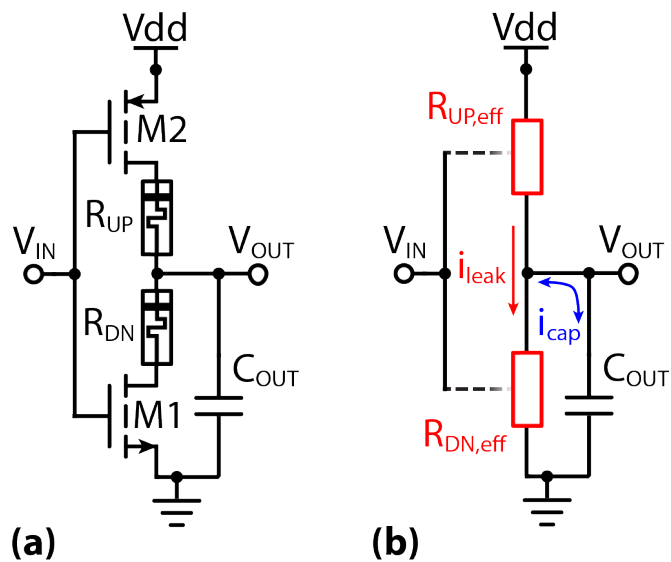

**Supplementary figure 2:** Simplifying assumption for power estimation analysis. (a) Analogue inverter topology. (b) Simplified version of (a). The combinations of $R_{UP}$–M2 and $R_{DN}$–M1 have been modelled as effective resistances $R_{UP,eff}$ and $R_{DN,eff}$ for fixed $V_{IN}$. Moreover, the key current components $i_{cap}$ and $i_{leak}$ are illustrated.

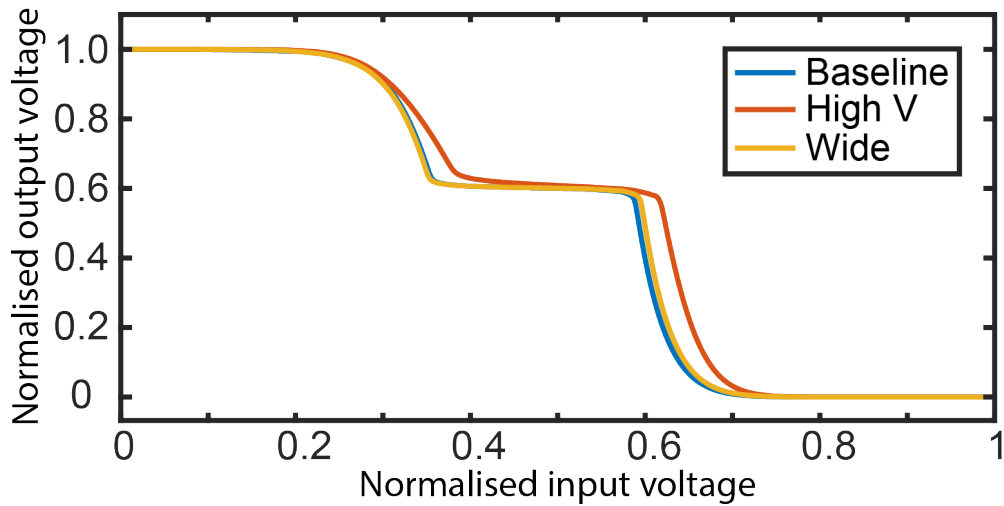

Supplementary figure 3: Trading off transistor sizings and power supply voltage against memristor resistive states. Three analogue inverter (Figure 1b) input/output transfer characteristics are shown, where two cases, 'High V' and 'Wide' are compared against a baseline case. In the 'High V' case the memristors operate at one order of magnitude lower resistive state vs. baseline and the power supply has been increased from 1.65V to 2.2V to compensate. In the 'Wide' case, the memristors also operate at one order of magnitude lower resistive state vs. baseline, but the inverter transistors have been increased tenfold to compensate. Interestingly, increasing the power supply leads to a slight shift in the location of the plateau, possibly because of the asymmetry between pMOS and nMOS characteristics. Notably, however, the power supply voltage can be chosen such that the plateau width remains the same. Detailed set-up parameters for each case can be found in Supplementary Table 1, including transistor specifications.

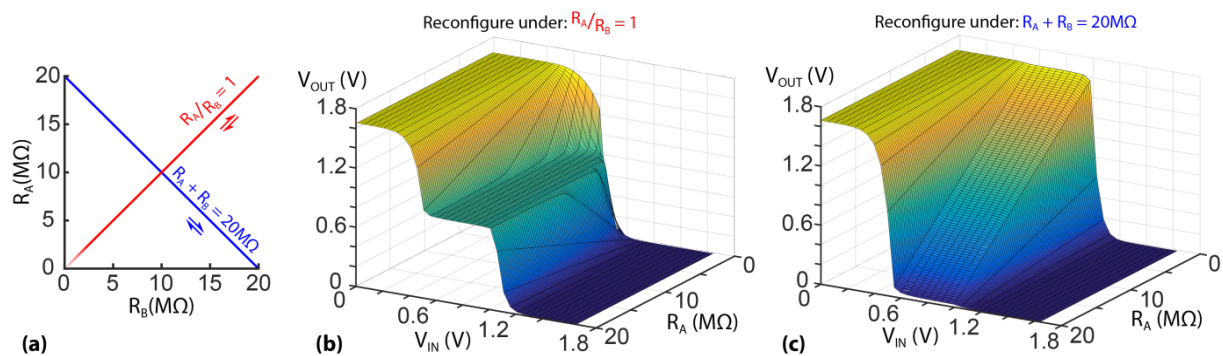

Supplementary figure 4: Reconfigurability of an analogue inverter akin to the one shown in Figure 1a. (a) Reconfigurability space demonstrated in (b) and (c), covering constant sum (=20MΩ) and constant ratio (=1) cases. (b) Constant ratio case. The plateau widens as $R_A$ (and consequently also $R_B$) increases. (c) Constant sum case. The altitude of the plateau decreases as $R_A$ increases (and consequently $R_B$ decreases). Transistor specifications as in baseline case of Supplementary Table 1.

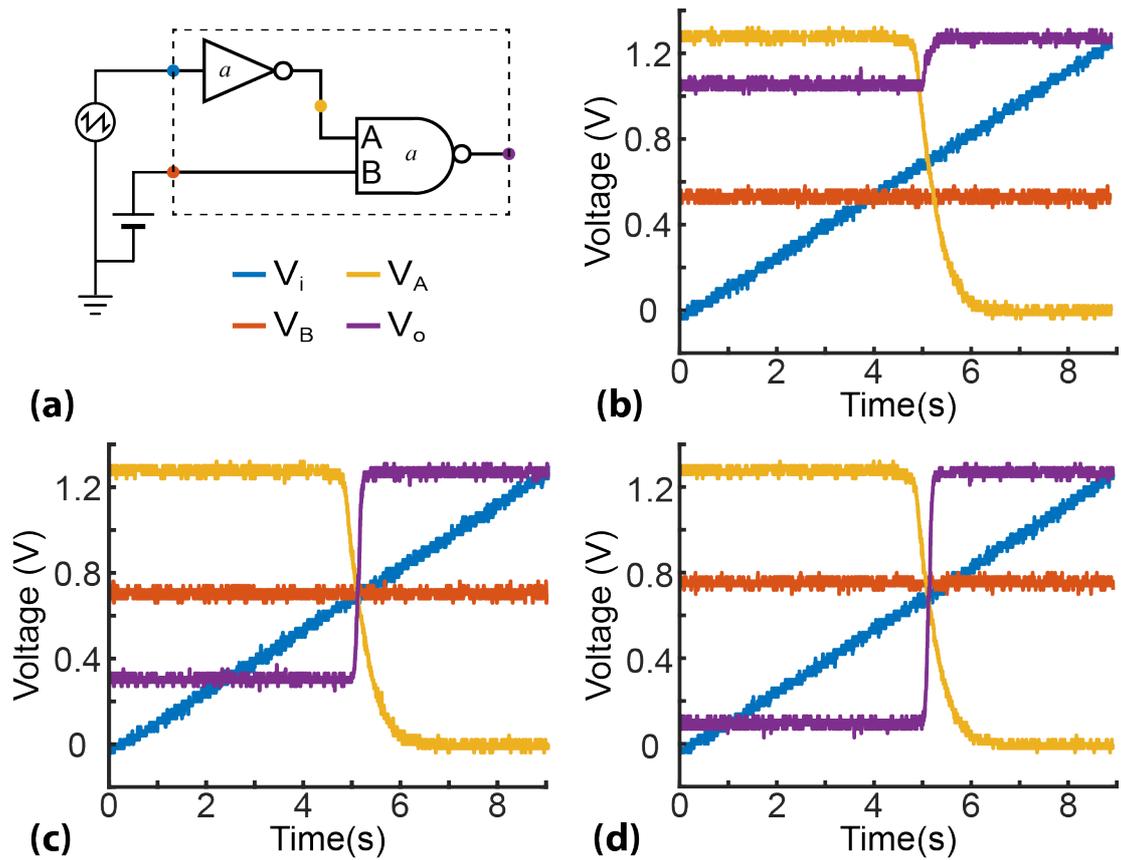

Supplementary figure 5: Example of analogue gates working together. (a) Topology tested: an analogue NAND gate with one input fed directly from a fixed voltage source and the other output fed from a saw-tooth source via an analogue inverter. (b-d) Three measured examples of the system input/output transfer characteristics, each taken for a different value of $V_B$. We notice that in (b), where $V_B$ is lowest the system always evaluates the function $\overline{A \cdot B} = \overline{\bar{\iota} \cdot B}$ as 'fairly true' ($V_O$ close to VDD). As the value of $V_B$ is increased, however, the system evaluates the same expression as less and less true. This occurs until $V_A$ drops below approx. 0.7V, in which case $V_O$ reaches VDD across all panels (b-d).

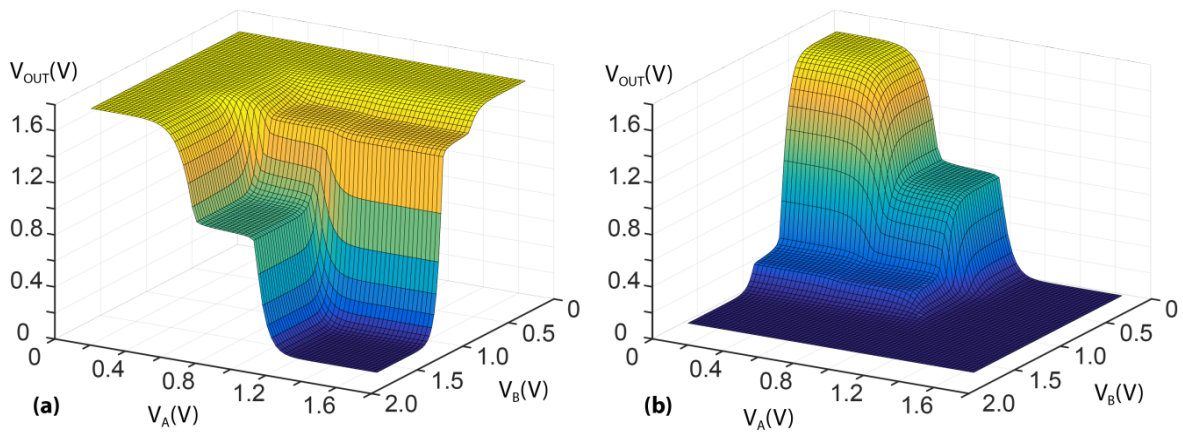

Supplementary figure 6: Demonstration of analogue NAND and NOR gates. (a) NAND gate input/output transfer characteristics. Schematic as shown in Figure 3a. $R_A$ = 3.5kΩ, $R_B$ = 0.5kΩ, $R_C$ = 4.0kΩ. (b) NOR gate input/output transfer characteristic. This gate is the exact dual of the NAND gate used for (a) (Exchange the power supplies to turn NAND into NOR and vice versa) and employs the same values of resistive states. pMOS/nMOS transistor specifications as in baseline case of Supplementary Table 1.

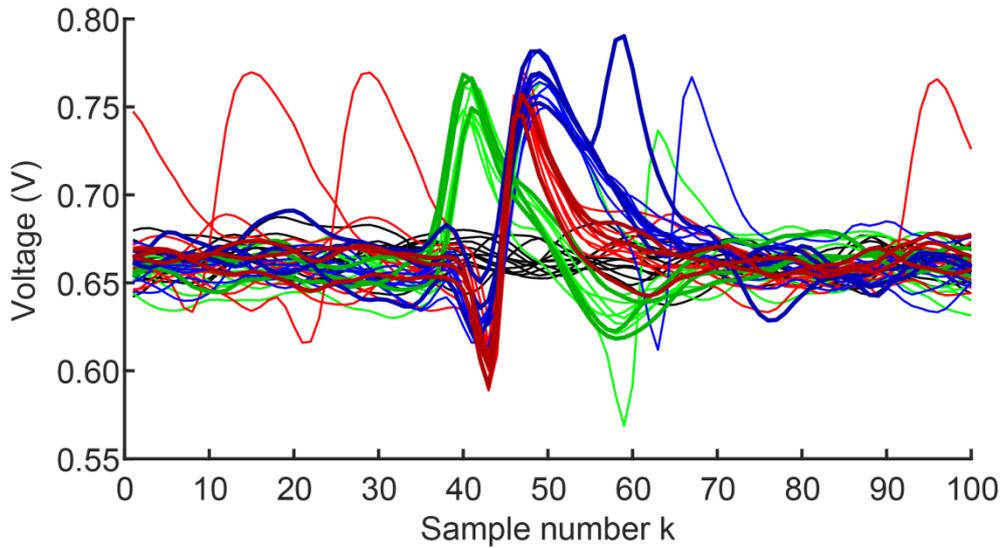

**Supplementary figure 7:** Dataset used for carrying out texel array experiment in Figure 4. Shown are all neural spike waveforms included in the dataset with colours indicating their class (same colour scheme as Figure 4). Spike waveforms chosen as inputs for the experiment in Figure 4 are shown as thicker, darker traces.

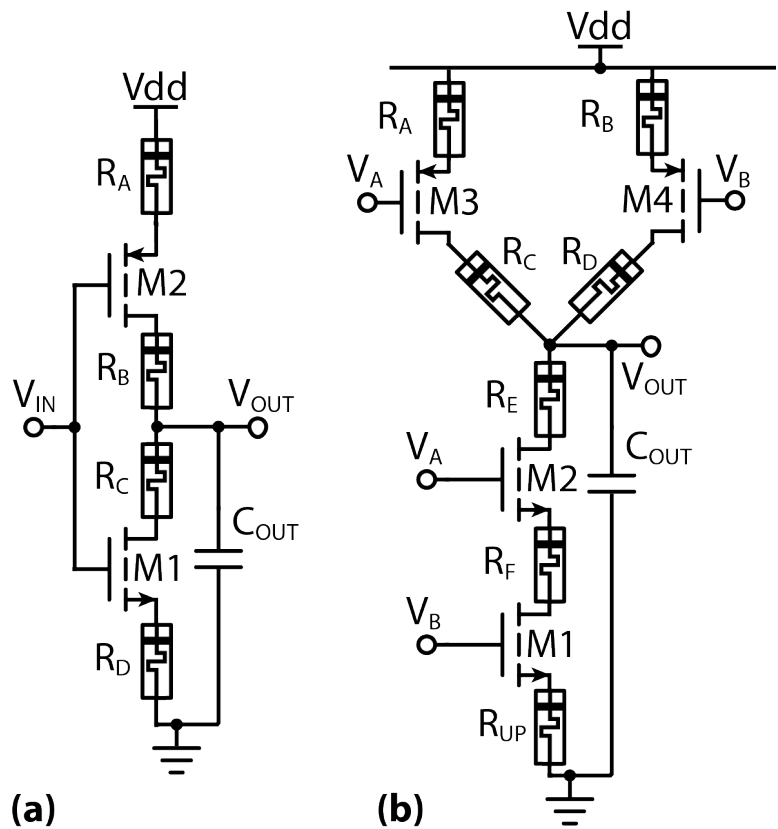

**(a)** **(b)**

**Supplementary figure 8:** Fully generalised memristor-based analogue gates. (a) Fully general analogue inverter. (b) Fully generalised NAND.

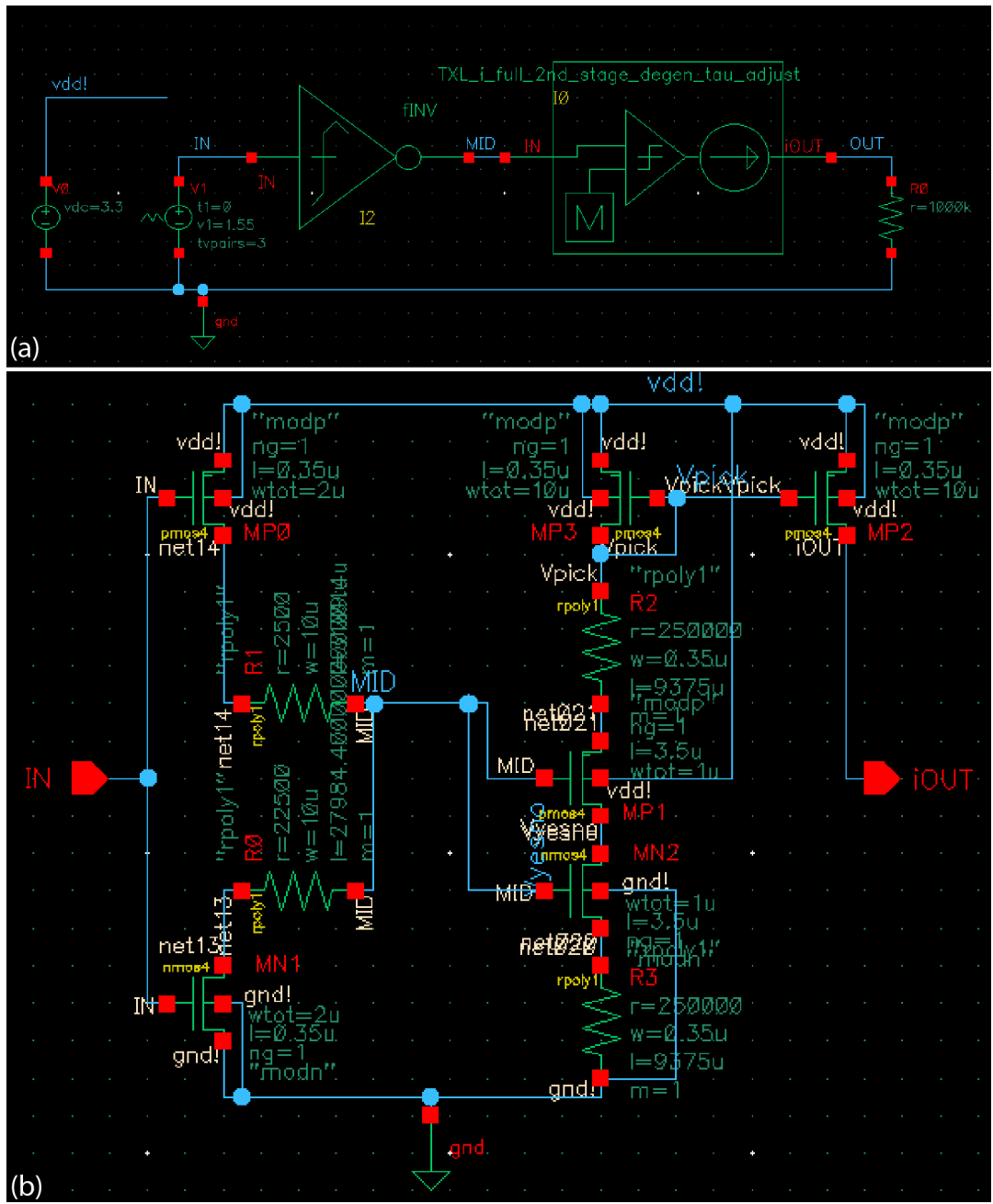

**Supplementary figure 9:** Schematics of texel power dissipation test bench (a) and texel circuitry (b) used to carry out power dissipation simulations. The driving inverter in (a) is similar to the inverter in (b), i.e. devices MP0, MN1, R0 and R1. In (b) memristors are represented by resistive elements. R0 and R1 are the memristors tuning the transfer characteristics of the texel whilst R2 and R3 are optionally implemented for tuning the sensitivity of the output current to input voltage and to act as current limiters. These memristors need not switch after fabrication and act as simple resistive loads.

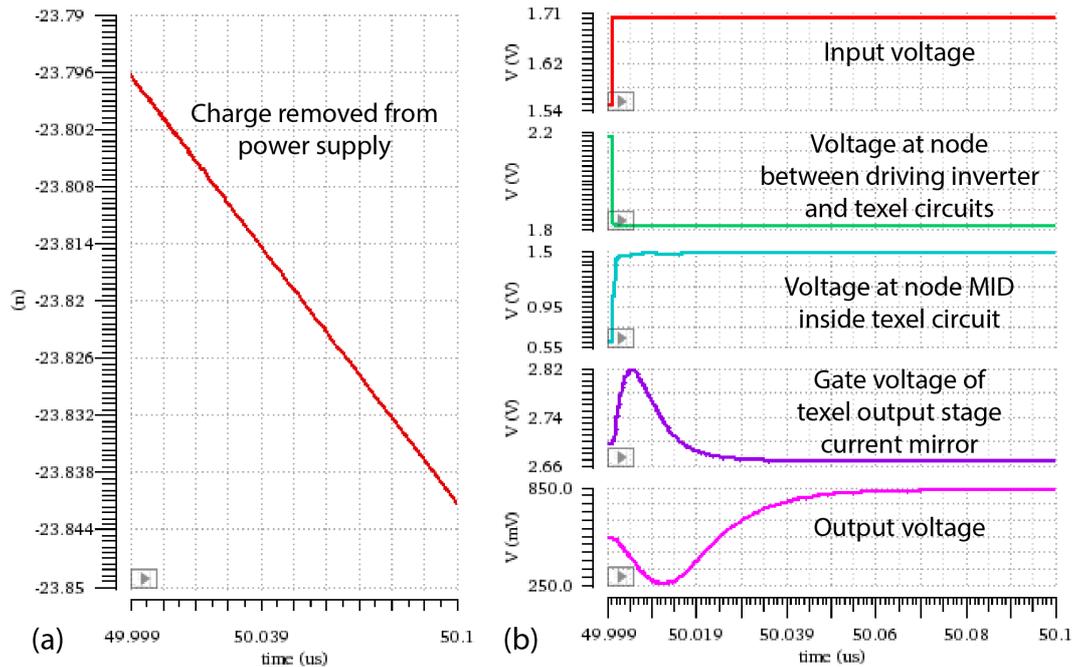

**Supplementary figure 10:** Charge dissipation of the test bench circuit in Supplementary Figure 9a for input signal transition from 1.55V to 1.70V. AMS 0.35 micron technology with power supply set to 3.3V. Approximately 46fC escape the power supply throughout the process, corresponding to toggling 37 minimum drive strength inverters as shown in Supplementary Figure 11. The design under study has not been optimised for power. (a) Charge dissipation through test circuit over time. (b) Selected voltage signal time evolutions from system input (red trace) to system output (green trace).

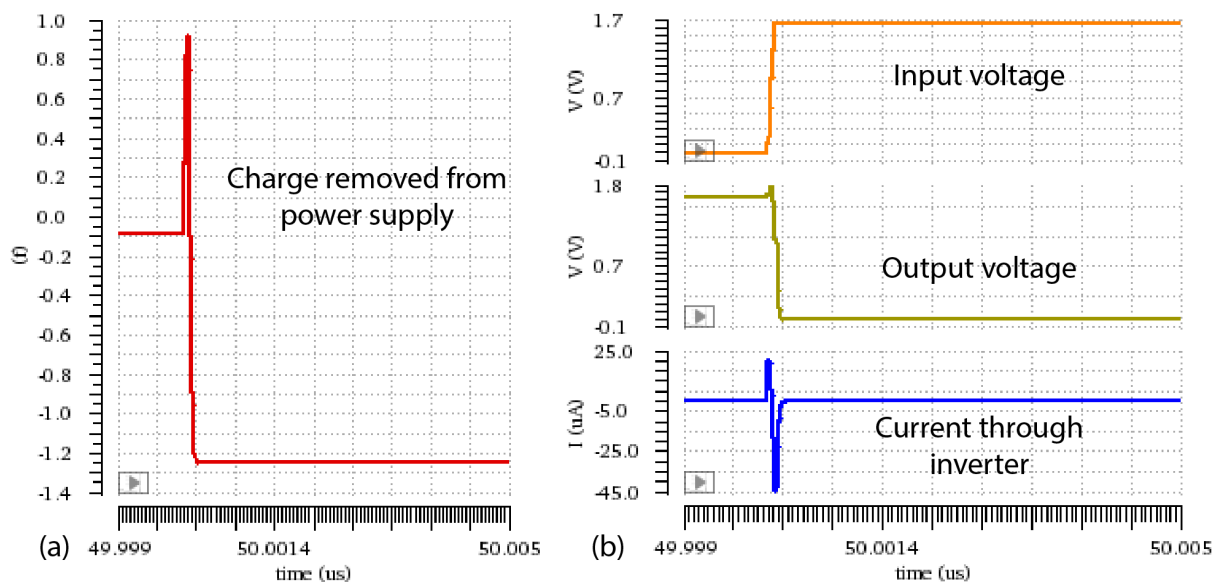

**Supplementary figure 11:** Charge dissipation of a minimum strength inverter in AMS 0.35 micron technology for a single input signal toggle. Approximately 1.25fC escape the power supply through the process. VDD = 1.65V. (a) Charge removed from power supply vs. time. (b) Input voltage signal. (c) Output voltage. (d) Current through the inverter.

## Supplementary tables:

|  | Baseline | Wide | High V |
|---|---|---|---|
| W2 (µm) | 12 | 120 | 12 |
| L2 (µm) | 3.5 | 3.5 | 3.5 |
| RUP (kΩ) | 4000 | 400 | 400 |
| RDN (kΩ) | 6000 | 600 | 600 |
| W1 (µm) | 1 | 10 | 1 |
| L1 (µm) | 3.5 | 3.5 | 3.5 |
| VDD (V) | 1.65 | 1.65 | 2.2 |

Supplementary table 1: Parameters used for simulations demonstrating the trade-off between power supply voltage, transistor sizing and memristive device resistive states. See Supplementary Figure 3.

|  | $R_{UP}$ (kΩ) | $R_{DN}$ (kΩ) |
|---|---|---|
| HH | 106 | 110 |
| HL | 106 | 20.9 |
| LH | 10.5 | 110 |
| LL | 10.5 | 11.5 |

Supplementary table 2: Resistive states of memristors, as measured under standard read-out voltage of 0.2V, used for the experiment in Figure 1.

|  | SPIKE CLASS | TXL1 | TXL2 | TXL3 | TXL4 | OUTPUT (1st run) | OUTPUT (2nd RUN) |
|---|---|---|---|---|---|---|---|
| Ideal | 3H | 0.7757 | 0.7657 | 0.7546 | 0.7443 | 0.03 | 0.06 |
| Rounded |  | 0.78 | 0.77 | 0.76 | 0.74 |  |  |
| Ideal | 3M | 0.7592 | 0.7510 | 0.7427 | 0.7350 | 0.08 | 0.16 |
| Rounded |  | 0.76 | 0.75 | 0.74 | 0.74 |  |  |
| Ideal | 3L | 0.7450 | 0.7397 | 0.7329 | 0.7266 | 0.25 | 0.39 |
| Rounded |  | 0.75 | 0.74 | 0.73 | 0.73 |  |  |
| Ideal | 2H | 0.7400 | 0.7270 | 0.7163 | 0.7071 | 1.00 | 1.19 |
| Rounded |  | 0.74 | 0.73 | 0.72 | 0.71 |  |  |
| Ideal | 2M | 0.7347 | 0.7231 | 0.7149 | 0.7094 | 0.99 | 1.23 |
| Rounded |  | 0.73 | 0.72 | 0.71 | 0.71 |  |  |
| Ideal | 2L | 0.7164 | 0.7058 | 0.6983 | 0.6923 | 0.58 | 0.83 |
| Rounded |  | 0.72 | 0.71 | 0.70 | 0.69 |  |  |
| Ideal | 1H | 0.7151 | 0.7055 | 0.6971 | 0.6904 |  |  |
| Rounded |  | 0.72 | 0.71 | 0.70 | 0.69 |  |  |
| Ideal | 1M | 0.7115 | 0.7014 | 0.6926 | 0.6863 | 0.36 | 0.62 |
| Rounded |  | 0.71 | 0.70 | 0.69 | 0.69 |  |  |
| Ideal | 1L | 0.6991 | 0.6888 | 0.6788 | 0.6705 | 0.14 | 0.28 |
| Rounded |  | 0.70 | 0.69 | 0.68 | 0.67 |  |  |

Supplementary table 3: Results for texel array experiment. Ideal (computed) and rounded values used as voltage inputs to the texel array elements (TXL1-4) are shown alongside the resulting output voltages at node VOUT for two repetitions of the experiment. All units are Volts.

|  | TXL1 | TXL2 | TXL3 | TXL4 |
|---|---|---|---|---|
| Before run | 19.5k | 14.8k | 12.7k | 10.6k |
| After run | 19.5k | 15.1k | 12.6k | 10.6k |

Supplementary table 4: Resistive states of memristors used for texel array experiment as measured before and after experimental run.

## Supplementary methods:

All experimental work carried out for the Supplementary material followed the same basic procedures and used the same instrumentation as explained in the methods section of the main text. All proof-of-concept level work e.g. Supplementary figures 3,4 and 6, was carried out using TSMC's MOSIS 0.35 micron technology with a power supply of 1.65V unless otherwise stated and LTSPICE. All quantitative analysis work, e.g. supplementary figures 10 and 11, was carried out using AMS' 0.35 C35 micron technology under a 3.3V power supply using Cadence. Component sizings are shown on the schematic of Supplementary Figure 9b. The read-out stage of the texel circuit in Supplementary Figure 9 was itself enhanced with memristors in order to drop overall power dissipation.